\documentclass[pre,preprint,endfloats]{revtex4}
\usepackage{amsfonts,amsmath,amssymb,mathrsfs}
\usepackage[usenames,dvips]{color}
\usepackage{graphicx}
\newcommand{\FF}{\ensuremath{\mathbf{F}}}
\newcommand{\XX}{\ensuremath{\mathbf{X}}}

\begin{document}
\title{Ensemble Inequivalence in Single Molecule Experiments}
\author{M. S\"uzen}
\author{M. Sega}
\affiliation{ Frankfurt Institute for Advanced Studies,
  Goethe-University, Ruth-Moufang-Str. 1, D-60438 Frankfurt am Main, Germany}
\author{C. Holm}
\email{holm@icp.uni-stuttgart.de}
\affiliation{Institute for Computational Physics, 
Stuttgart University, Pfaffenwaldring 27, 70569 Stuttgart, Germany}

\begin{abstract}
  In bulk systems the calculation of the main thermodynamic quantities
  leads to the same expectation values in the thermodynamic limit,
  regardless of the choice of the statistical ensemble.  Single linear
  molecules can be still regarded as statistical systems, where the
  thermodynamic limit is represented by infinitely long chains. The
  question of equivalence between different ensembles is not at all
  obvious and has been addressed in the literature, with sometimes
  contradicting conclusions. We address this problem by studying the
  scaling properties of the ensemble difference for two different
  chain models, as a function of the degree of polymerization.  By
  characterizing the scaling behavior of the difference between the
  isotensional (Gibbs) and isometric (Helmholtz) ensembles in the
  transition from the low-stretching to the high-stretching regime, we
  show that ensemble equivalence cannot be reached for macroscopic
  chains in the low force regime, and we characterize the transition from
  the inequivalence to the equivalence regime.
\end{abstract} 

\maketitle

\section{Introduction}
Among the technical developments that influenced the research in
molecular biophysics during the last two decades, the possibility of
manipulating single molecules by means of different techniques has
certainly played a pivotal role, and has been proved to be an
invaluable tool to gain insight into the structural properties and
function of macromolecules involved in many biological
processes\cite{ritort06a}. The techniques devised to manipulate single
molecules are collectively known as single molecule experiments (SMEs)
\cite{merkel01a,strick03a,ritort06a} and include, among others,
approaches based on atomic force microscopy
\cite{binning86a,rief97a,oesterhelt99a,fisher00a,engel00a}, optical
tweezers \cite{wang97a,liphardt01a,wen07a,moffitt08a}, fluorescence
detection of F\"orster resonance energy transfer \cite{lipman03a},
elongational flow \cite{perkins97a,neumann99a,weiss00a} and magnetic
tweezers \cite{smith92a,strick96a,besteman07a}.

It is widely known that equilibrium thermodynamics can be recovered
from a statistical mechanics description in the limit of infinite
system size, (the thermodynamic limit) and that, except when in
proximity of a phase transition, all statistical ensembles provide the
same average values for the observables of interest. In other words,
in the thermodynamic limit different ensembles become equivalent.

In case of SMEs, on the contrary, the outcome of a measure explicitly
depends on the control parameters, that is, on the choice of which
quantities are kept constant and which ones are allowed to vary.  For
this reason, the efforts in providing satisfying theoretical
descriptions of small, out of equilibrium systems have been
intensified during the last
decade.
In particular, the validation of the hypothesis made by
Flory\cite{flory89a}, that ensemble equivalence for a SME on a linear
polymer should be obtained in the limit of infinite chain length, has
been often subject of
investigations\cite{ritort06a,keller03a,neumann03a}. Since one of the
basic means of extracting information regarding a linear polymer in a
SME is to analyze the molecular force--extension curve
(FEC)\cite{ritort06a}, it is natural to introduce two different
conjugate ensembles, namely, the Helmholtz, or isometric ensemble, and
the Gibbs, or isotensional ensemble. In the isometric ensemble the
position of the chain ends is employed as a control parameter, fixing
also the end-to-end distance, while in the isotensional case the
control parameter is represented by the force applied on one loose
end.  The conditions under which the equivalence between these two
ensembles can be obtained have been investigated in many works, both
from a theoretical point of view
\cite{guth34a,pincus76a,perchak82a,weiner82a,neumann85a,neumann86a,weiner87a,winkler92a,kreuzer01a,makarov02a,winkler03a,keller03a,hanke05a,sinha05a,vanderstraeten06a,skvortsov06a,hanke06a,ranjith05a,rubi06a}
and by means of computer
simulations\cite{webman81a,weiner81a,weiner85a,berman85b,guyer85a,cifra95a,titantah99a,zemanova05a,guardiani06a,linna08a}.
Some of the authors concluded that ensembles are equivalent in the
infinitely long chain limit for a Gaussian
chain,\cite{winkler92a,titantah99a} as well as for a generic chain
\cite{keller03a} and toy lattice chain models
\cite{keller03a,sinha05a,vanderstraeten06a}. Some other works states
that ensembles are not equivalent in the thermodynamic limit for a
single chain\cite{neumann85a,berman85b,guyer85a,neumann86a}.

The present paper addresses the problem of ensemble equivalence of
Gaussian chains from the point of view of computer simulations,
employing coarse-grained Langevin dynamics simulations.  In Section 2,
we discuss the basic theoretical background and the appropriate way of
quantifying the ensemble equivalence.  After we describe the employed
simulation techniques in Section 3, we present the analysis of the
force-extension curves in the Helmholtz and Gibbs ensembles for
Gaussian chains with zero and non-zero average bond vector in Section
4. There we also discuss the analysis of the scaling behavior of the
ensemble difference with respect to the chain length, confirming the
prediction of Neumann\cite{neumann03a} on the ensemble inequivalence
in the limit of vanishing applied forces, before we conclude in
Section 5.

\section{Conjugate ensembles and their equivalence in a SME}
In statistical mechanics the connection with thermodynamics is realized by defining the
thermodynamic potentials from the partition functions, in the limit that
every extensive control parameter is going to infinity.
As a general result, given a statistical ensemble and a control parameter, it is possible to
construct a conjugate ensemble using as a conjugate control parameter the derivative of 
the ensemble's thermodynamic potential. It is generally assumed that in the
thermodynamic limit, two conjugate ensembles should be equivalent,
namely, they should provide the same expectation value for the
thermodynamic quantities\cite{huang1987}. Let us take as an example the case 
of the canonical ensemble, whose partition 
function will be denoted by $Q(N,V,T)$, and whose thermodynamic potential is
the Helmholtz energy $A$, defined as $$-\beta A =
\lim_{(N,V)\to\infty} \ln Q(N,V,T).$$ Here $\beta = 1/k_{B} T $, where 
$k_{B}$ is Boltzmann's constant. The main thermodynamic
quantities can be then derived by computing the derivatives of the
thermodynamic potentials with respect to their parameters. Given any generic
ensemble, characterized by its thermodynamic potential $P(Y_1,Y_2\,\ldots)$, it is
possible to generate a conjugate ensemble by choosing  any quantity, 
$Y_i$, and applying a Legendre transform which involves the
pair of conjugate variables $Y_i$ and $\partial P / \partial Y_i$\cite{callen,chandler,greiner}.
In the example of the canonical ensemble, by choosing the volume $V$ and the pressure $p =
-\partial A / \partial V$ as a conjugate pair, the isothermal-isobaric ensemble ($NpT$) is generated. 

The Legendre transform applied to the Helmholtz 
energy yields the thermodynamic potential of the isothermal-isobaric
ensemble, namely, the Gibbs energy $G(N,p,T) = A + pV$.  The
conjugate partition function, $\mathscr{Z}(N,p,T) $ can be written in
a natural way as a weighted sum of the canonical partition functions
\begin{equation}
     \mathscr{Z}(N,p,T)   =   \int \mathrm{d}V Q(N,V,T)  \exp(-\beta p V).
\label{eq:laplace}
 \end{equation}
Consequently, the ensemble average $\left\langle \mathscr{A}\right\rangle$ of any observable 
$\mathscr{A}(V)$ can be expressed in the conjugate ensemble as
\begin{equation}
     \left\langle\mathscr{A}\right\rangle  =  \mathscr{Z}^{-1} \int \mathrm{d}V Q(N,V,T)  \exp(-\beta p V) \mathscr{A}(V).
 \end{equation}

 The ensemble equivalence problem can then be stated as follow: \emph{can
 the thermodynamic potential of the conjugated ensemble be obtained
 from the thermodynamic limit of the conjugate partition function?}
 If this is true, the thermodynamic quantities computed in either
 ensembles will lead to the same expectation values. Conversely,
 if at a given thermodynamic point a generic function of state
 assumes different values, depending on the ensemble in which it is
 computed, then the ensembles are inequivalent.

\begin{figure}[ptb]
\begin{center}
\includegraphics[width=0.5\columnwidth]{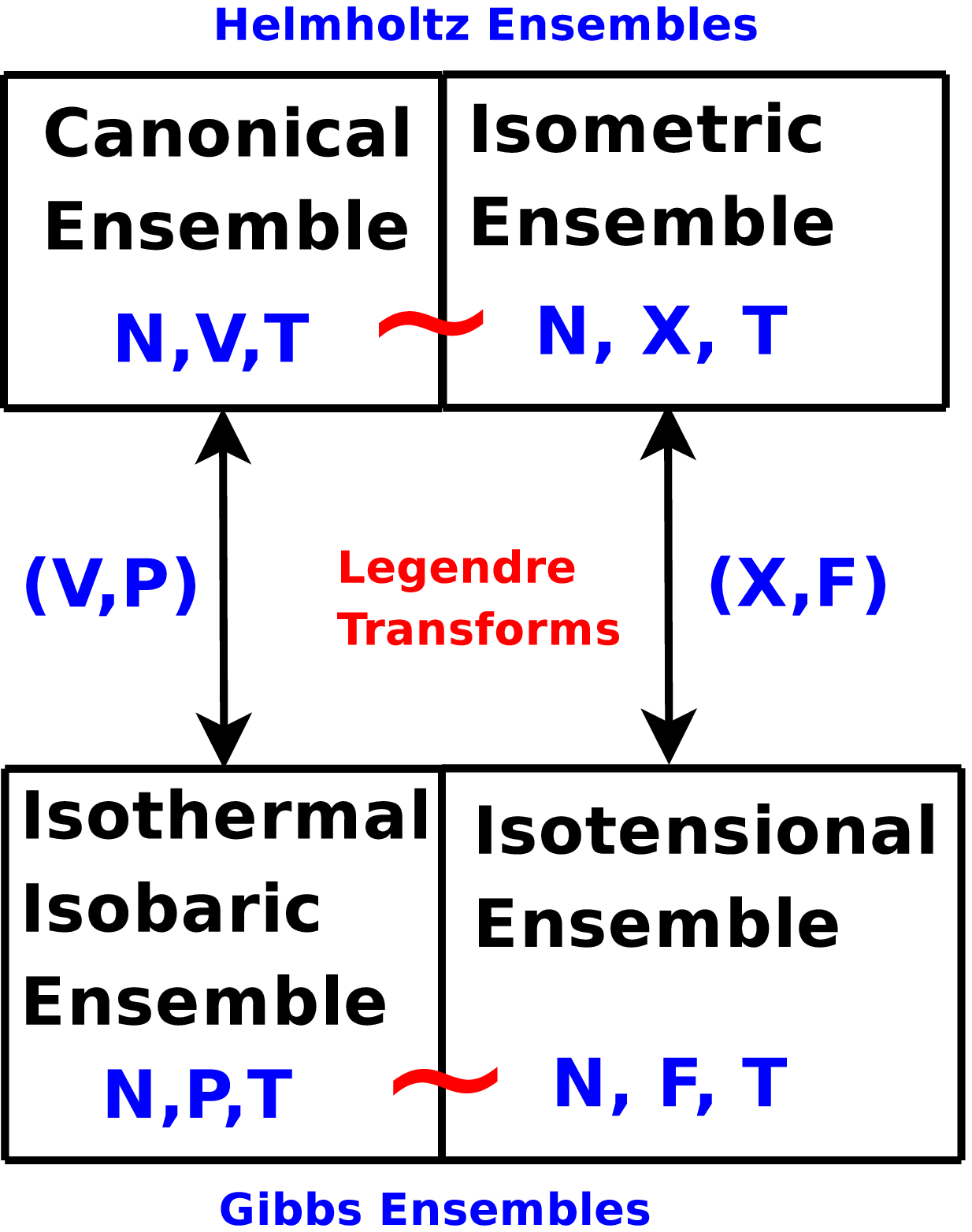}
\end{center}
\caption{The sketch of analogies between conjugate chain ensembles and standard statistical mechanical ensembles.}
\label{ensemble_analog}
\end{figure}

As pointed out first by Flory\cite{flory89a},  single polymers can be regarded as
statistical ensembles, and their thermodynamical limit can be realized by infinitely long chains.
Among the interesting control parameters that can be employed to characterize the statistical
ensemble of a molecule is the molecular end-to-end (displacement) vector $\XX$. Following
the convention of Neumann\cite{neumann03a} we will call isometric the statistical ensemble
generated by keeping this vector constant. Obviously, $\XX$ \emph{plays no privileged role with
respect to other observables}, and other choices for the control parameter (like the end-to-end
distance) are possible. Here we will restrict our analysis to the end-to-end vector $\XX$ (as a control parameter),
because we consider it to be relevant for experiments. For example, the end-to-end distance
$\left|\XX\right|$ would be the control parameter in an experiment for which the molecular
ends are kept at constant distance, but where the end-to-end vector is free to rotate. This
ensemble, although perfectly licit, would obviously be difficult to be realized experimentally.

Keeping the other control parameters (temperature, number of particles, $\ldots$)
fixed, the work done on the system by displacing the chain end by an amount
$\mathrm{d}\XX$ is $\mathrm{d}U = \FF\cdot \mathrm{d}\XX$. As a consequence, the
force $\FF$ acting on one end is equal to the derivative with respect to $\XX$ of
the thermodynamic potential. Therefore, $\FF$ is the variable conjugate to $\XX$.
Note that since a  conjugate pair  ($ Y_i $, $\partial P/\partial Y_i $) is uniquely
determined by the ensemble and a control parameter, in the isometric ensemble
defined above the force $\FF$ is the only possible conjugate variable to $\XX$.
This leads, via a Laplace transform analogous to (\ref{eq:laplace}) to the
definition of the ensemble conjugate to the isometric one, namely, the isotensional
one. In the isotensional ensemble a constant force is applied on one molecular
ending. A schematic representation of the formal analogy between the isometric,
isotensional, canonical and isothermal-isobaric ensembles is presented in
Fig.\ref{ensemble_analog}.  

In experiments in which single molecules are directly manipulated, as is the case for AFM
or optical tweezers, the manipulating device is basically acting on the terminal part of the
molecule, and depending on the strength of the interaction of the tool with the molecule,
the former can be employed either to hold firmly one molecular end or, to measure the force
which is acting on it, while the other end is attached to some rigid support\cite{keller03a}.
Both the isometric and the isotensional ensembles can then be realized by changing the force
constant of the cantilever in the case of AFM or spring constants in the case of optical
tweezers experiments\cite{kreuzer01b}, where the use of high force constants will lead to a
sampling of the isometric ensemble. It is therefore of particular interest,
both from a fundamental
 and from a practical point of view, to understand if these two ensembles are equivalent.
 This knowledge
can tell, e.g., whether measurements performed in the two ensembles are comparable, or, which
kind of theoretical model has to be employed in order to fit experimental data, which is a
key step to extract information from SMEs\cite{oesterhelt04a,marszalek02a}.

The term ``ensemble equivalence''  indicates,   as it is stated in textbooks on
statistical mechanics\cite{huang1987,hill1957,huang1987,greiner}, that in the
thermodynamic limit the partition functions of two ensembles become indistinguishable.
As a result, if two ensembles are equivalent, the expectation values of \emph{any}
observable measured in the two ensembles have to coincide.

Ensemble equivalence requires that a calculation of \emph{any observable} in the thermodynamic
limit should yield the \emph{same expectation value} for both ensembles\cite{touchette04a}, and this is 
a sound definition of equivalence from the practical point of view.  Surely it is possible
to find some specific observable  that has the same expectation values in the two
ensembles, but this is by no means a proof of ensemble equivalence, unless one checks this for 
every observable. In this sense the negative
proof is an easier task: if one manages to find an observable which behaves differently in
the two ensembles, their inequivalence is proven. We will show that the expectation values 
for the end-to-end distance $X=\left|\XX\right|$ sampled in the isometric and isotensional
ensembles converge only by choosing a thermodynamic point characterized by high values
of $\left<\FF\right>$, while they do not converge in the weak forces regime. This will prove
that in the part of the phase diagram characterized by weak applied forces the two ensembles
are not equivalent.

To summarize, the basics steps taken to define the isotensional and isometric ensembles and to check their inequivalence are the following
\begin{enumerate}
\item The end-to-end vector $\XX$ is chosen as a control parameter. 
      We denote the ensemble generated by keeping this parameter
      fixed as \emph{isometric}.
\item The variable conjugate to $\XX$ is $\FF$ (since the work
      done on the system is $\mathrm{d}U=\FF\cdot\mathrm{d}\XX$).
      We denote the ensemble generated by keeping $\FF$ fixed 
      as \emph{isotensional}. It is realized by applying a constant 
      force on one molecular end.
\item The value of the applied force (in the isotensional ensemble) or
      of the average force $\langle\FF\rangle$ (in the isometric ensemble) identifies
      a thermodynamic point of the system.
\item The extension $X=\left|\XX\right|$ is chosen as the observable to be compared in the
      two ensembles. This observable assumes constant values in the isometric ensemble and 
      fluctuates in the isotensional one.
\item The relative difference between the average values of $\left|\XX\right|$ is 
      sampled in the two ensembles at different thermodynamic points, that is, different values of 
      $\langle\FF\rangle$, and its scaling is investigated in the limit of long chains.
\end{enumerate}
In a system which does not exhibit any directional preference, due to symmetry reasons the 
average force $\langle\FF\rangle$ points along $\XX$. For that, a thermodynamic point can be uniquely identified by the
modulus of the average force $\left|  \langle\FF\rangle  \right|$ or, equivalently, by the average projected force defined as
$\langle F\rangle \equiv \langle \FF \cdot \XX / X\rangle$, since it holds that  $\langle F \rangle = |\langle \FF \rangle|$.
Therefore, checking the convergence of $\left|\XX\right|$ at different thermodynamic points is equivalent to check the 
convergence of the force-extension curves  $\langle F\rangle  (X)$ and $F(\langle X\rangle)$.

In case of ensemble equivalence, these two graphs should, in the limit of infinitely long
chains, become indistinguishable. Note, however, that some care has to be taken, about the
precise meaning of this statement, as will be discussed in the next section.

\section{Ensemble equivalence for Gaussian chains\label{quantens}}
In general terms, if the probability density distribution for the
end-to-end vector at equilibrium in the free case is denoted
by $\mathscr{P}(\XX)$, then the partition functions for the isometric and isotensional cases
can be written, respectively, as
\begin{eqnarray}    
    Z_{\XX}& =& \mathscr{P}(\XX)  \\
    Z_{\FF}& =& \int \mathrm{d}\XX \mathscr{P}(\XX) \exp(-\beta \FF\cdot \XX),
\end{eqnarray}  
where the formal analogy with the canonical and its conjugate
isothermal-isobaric ensemble is evident.  Often, as is the case with
computer simulations and different theoretical approaches, one needs
to refer to a specific molecular model, which will explicitly
determine the form of the probability density $\mathscr{P}(\XX)$. In
case of long linear chains, however, the probability density is very well 
approximated by a Gaussian
distribution\cite{doi86a}
\begin{equation}
      \mathscr{P}(\XX) = b^{3} \pi^{-\frac{3}{2}} \exp(-b^{2} X^{2}), \label{gauss_dist}
\end{equation}
where the characteristic length $b^{-1}$, is proportional to the root-mean-square
end-to-end distance $\sqrt{\left\langle |\mathbf{X}|^2
  \right\rangle}_0 = \sqrt{3/2}/b$ and to the mean end-to-end distance
$\langle X \rangle_0 = 2/(b\sqrt{\pi})$ (the notation $\langle
\ldots\rangle_0$ denotes an ensemble average in the free case,
i.e. with no applied force). This distribution can be recovered as a
limiting case for a wide class of different models representing
polymers with discrete units. This is the case, for example, for a
freely jointed chain of $N$ elements of length $a$, whose end-to-end
distribution is well approximated by the Gaussian distribution when $N
a \gg X$. In this case, $b^{2}=3 / (2Na^{2})$ and the total contour
length $L$ is $L=Na$. Moreover, the Gaussian chain model can be easily simulated
by means of a series pointlike particles connected by harmonic springs.

Given this explicit form for the partition function, it is possible 
to make some analytical prediction on the model. As Neumann noted in a critical analysis on the
interpretation of stretching experiments\cite{neumann03a}, from the
partition functions of
the Gaussian chain it is possible to derive Hookean-like relations
between the force and end-to-end vectors. This is easily seen by computing
the derivative of the free energy with respect to the end-to-end
vector in the isometric case
\begin{equation}   
  \left\langle \FF \right\rangle=  \beta^{-1}
  \frac{ \partial}{\partial \XX} \ln Z_{\XX}  = 
  2 b^{2} \XX / \beta   \label{FEC_X}
\end{equation}
and computing the average end-to-end vector $\left\langle \XX \right\rangle$ in the isotensional case
\begin{equation}   
  \FF= 2 b^{2} \left\langle \XX \right\rangle / \beta.
  \label{FEC_F}
\end{equation}
This analogy has usually led to the erroneous interpretation of an
equivalence between the isometric and isotensional ensembles, which
moreover seems to hold for every chain length, and not just in the
thermodynamic limit of infinite chain length.

Disregarding the fact that this equivalence is not
true for other observables has been often a source of misinterpretations,
for example when theoretical force-extension curves that predict
zero extension at zero applied force are employed
(see, \emph{e.g.},\cite{perkins97a,marko95a,li06a,doi78a}) either
to fit experimental data or to build theoretical models. In these
cases \emph{ad-hoc} corrections or hypotheses have been admittedly
introduced. This could have been avoided by employing the right
ensemble and, consequently, non-vanishing extensions at zero applied
force.

In fact, it is clear that these relations cannot hold for other
observables such as, \emph{e.g.}, the end-to-end distance $X$
(remember that in order to attain ensemble equivalence every
observable should converge to the same limiting expectation value).
This can be inferred  from the well known fact that in the free
case the average end-to-end distance $\langle X \rangle_0$ does not
vanish, whereas Eq.(\ref{FEC_X}) yields zero force at zero distance.
Therefore, by choosing observables other than  $\mathbf{X}$,
the convergence of their expectation values is not a trivial matter anymore.
A direct evaluation\cite{neumann03a} of $\left\langle X \right\rangle$ in
the isotensional ensemble, in fact leads to
\begin{equation} \label{fluct_x} 
\langle X \rangle   =  \frac{ \langle X \rangle_{0}}{2} \left[ e^{-\nu^2 / 4} + (\nu + 2/ \nu ) \int_{0}^{\nu/2} e^{-t^2} \mathrm{d}t  \right],
\end{equation}
where $\nu=\beta F/b$.  This relation can be usefully approximated,
for small external forces, as $ \langle X \rangle = \langle X
\rangle_{0} \left[ 1 + \frac{1}{12} \nu^{2} + \mathcal{O}(\nu^{4})
\right] $, demonstrating a nonzero extension in the weak force limit
in fact for all chain lengths. Therefore, the $F$ vs $X$ force-extension
curves measured in the two different ensembles can not coincide.
The question of ensemble equivalence has then to be investigated
more carefully, by examining the scaling behavior (\emph{i.e}
the finite size effects) of an observable that quantifies the difference
between the expectation values of $X$, when measured in different ensembles.

\begin{figure}[ptb]
\begin{center}
\includegraphics[width=0.7\columnwidth]{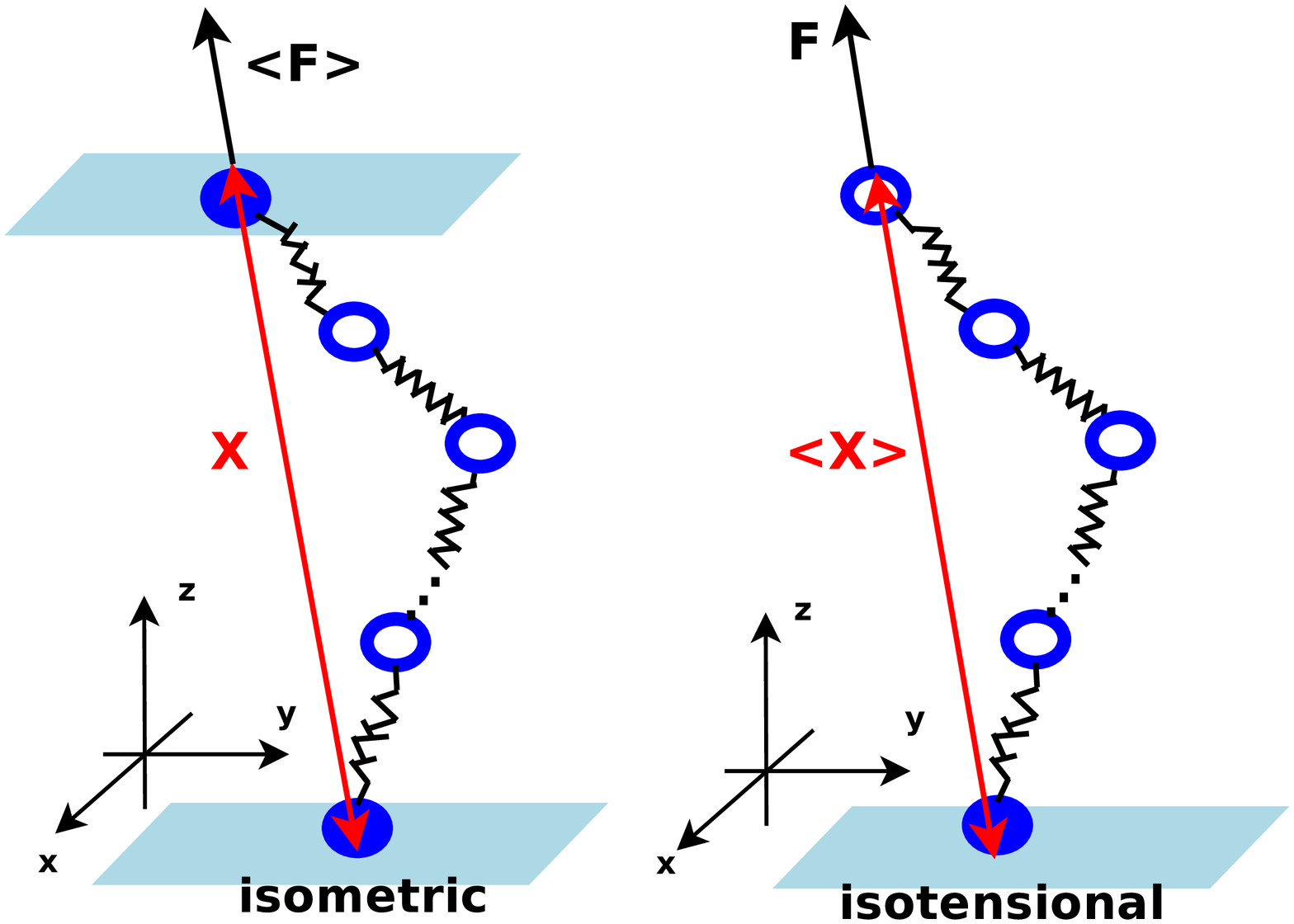}
\end{center}
\caption{Sketch of the chains in the isometric and isotensional ensembles. 
}
\label{gauss_chain_ensembles}
\end{figure}

In recent investigations the fundamental difference in meaning between
Eqs. (\ref{FEC_X}) and (\ref{FEC_F}) was not appreciated resulting in
a general consensus that in the thermodynamic limit, equivalence
between the two ensemble is obtained. In particular, employing
techniques like renormalization group theory\cite{titantah99a},
maximum entropy approach\cite{winkler92a}, and standard analytical
approaches \cite{keller03a,sinha05a}, the authors conceived the idea
that the isotensional and the isometric ensemble are actually
equivalent in the usual thermodynamic limit.  Recently, however,
Neumann pointed out\cite{neumann03a} a particular feature of the
statistical mechanics of the single chain, namely, that ensemble
equivalence cannot be obtained for small values of the external
force. In other words, there is theoretical evidence that by choosing
an appropriately small external force, the ensemble difference for any
(finite, but arbitrarily large) chain length can be maintained
constant. To our knowledge, this subtle point in the investigation of
the equivalence has never been tested by means of computer simulation,
and a part of this paper is devoted to clarify it, in particular by
analyzing the scaling behavior of the ensemble difference as a
function of the applied force regime.

Let us turn now to the problem of evaluating ensemble difference and its scaling.
In order to make any statements about equivalence in the thermodynamic
limit, there is need for an observable which quantifies how much 
two ensembles differ. Given a point $(F^*,X^*)$ on the graph of the
isotensional force-extension curve $F(\langle X\rangle)$, the measure
of the difference between two ensembles is then defined as
\begin{equation}
\Delta = \frac{X^* - X_{mp}}{X^*},\label{difference}
\end{equation}
where $X_{mp}$ is the value of the extension in the isometric ensemble
that solves the equation $\langle F \rangle (X^*) = F^*$. (The
notation stems from the fact that this extension actually coincides
with the maximum in the probability distribution function of the
isotensional ensemble\cite{keller03a}) 
This definition (which of course is not restricted to Gaussian chains) 
plays a crucial role, since the difference between 
the two ensembles \emph{should not depend on a reparametrization} of the
control parameter. So, while $\Delta^\prime(X) = X^* - X_{mp}$ could
in principle seem to be a reasonable alternative, it is easily seen
that using the same functional form $\Delta^{\prime}(\xi) = \xi^* -
\xi_{mp}$ for the relative extension $\xi=X/N$ would lead to completely
different results. This does not happen using the relative
measure (\ref{difference}), which behaves correctly under reparametrization.

\begin{figure}[ptb]
\begin{center}
\includegraphics[width=0.60\columnwidth]{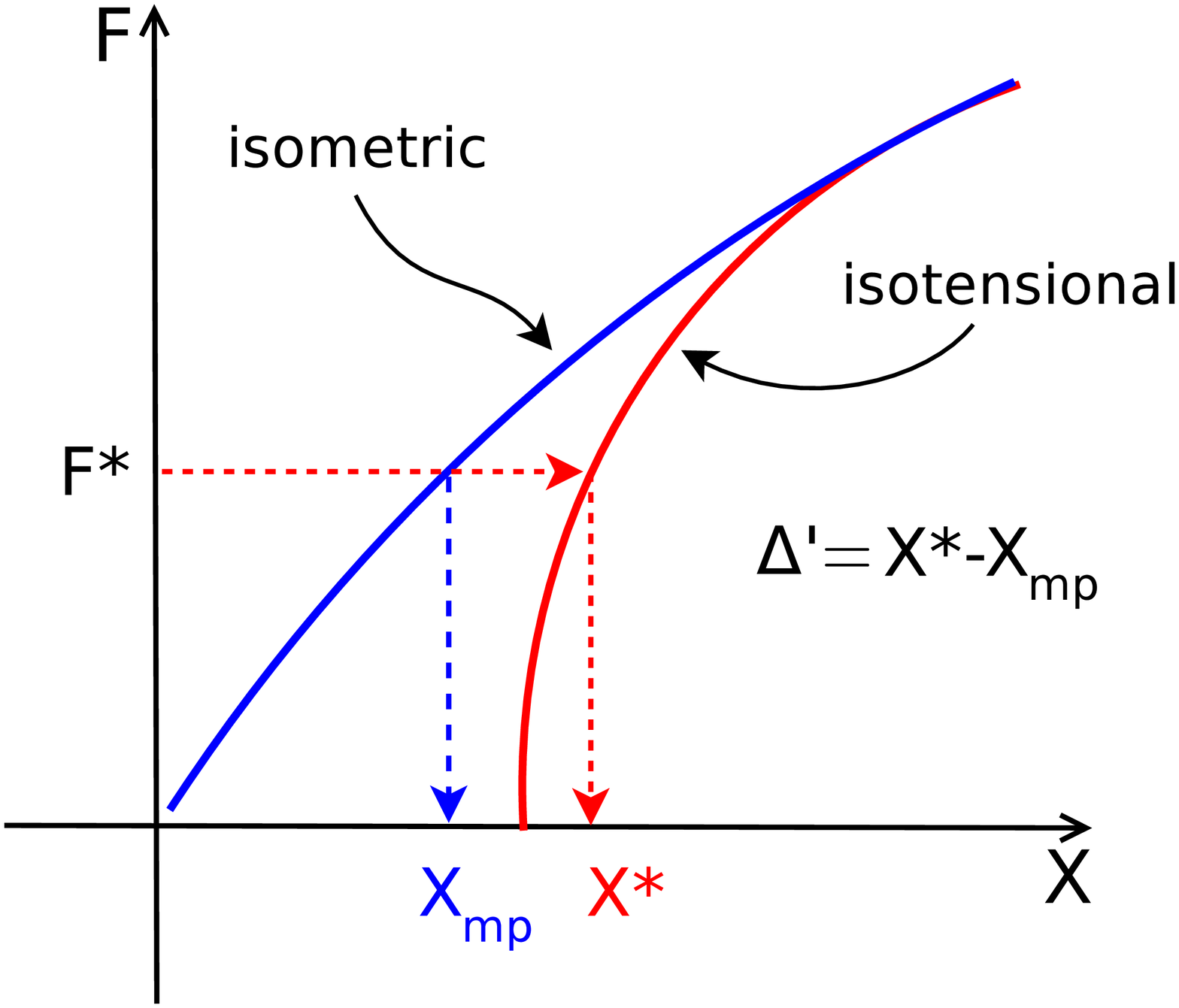}
\end{center}
\caption{Definition of ensemble difference $\Delta^\prime$ between
  isotensional and isometric ensembles for a given value $F^*$ of the
  force applied in the isotensional ensemble.\label{def_ens_diff} }
\end{figure}

A schematic view of the
identification of $X^*$ and $X_{mp}$ from the graphs of the
isotensional and isometric force-extension curves is given in
Fig.\ref{def_ens_diff}. 

The correct definition of the ensemble difference shows (at least for the Gaussian
chain) that in the moderate or strong stretching regime the ensemble
difference indeed goes to zero when the number of monomers tends to
infinity, i.e., that ensemble equivalence is attained.  In particular,
if the chain end-to-end distance is supposed to scale linearly with
the number of monomers in the over-stretched regime, the ensemble
difference should scale\cite{keller03a} like $\Delta \sim (N-1)^{-1}$.
However, the behavior of the ensemble difference at low stretching
regimes is markedly different, inasmuch as in this regime both $X^*$
and $X_{mp}$ scale as $\sqrt{N-1}$, and therefore the ensemble
difference does not scale with system size, and remains constant. For
any given chain length (i.e. for a macroscopic, though not strictly
infinite system) it is then possible to find a small enough force for
which the ensemble difference does not vanish and, moreover does not
decrease appreciably when increasing the chain length. The ensemble
difference in the free case $\Delta_0$, as a limiting case of zero
forces, can be written as $$ \Delta_0 = 1 - \frac{X_{mp,0}}{X_0^*}.$$
Since for Gaussian chains $X_{mp,0}=0$, the ensemble difference takes
the limiting value of $\Delta_0=1$, which for long enough chains is a
model-independent result, as long as the Gaussian approximation is
valid. 
Summarizing, the analytical argument of Neumann
demonstrates that, in principle,  if one chooses a thermodynamical
point characterized by a small enough force, then the expectation
values for $X$, measured in the two ensembles will not converge, therefore
exhibiting ensemble inequivalence in the vanishing force
regime. However, the presence of two simultaneous limiting procedures
(\emph{i.e.} of vanishing forces and of diverging chain lengths)
introduces some difficulties in its interpretation, and poses the
serious question wheter the set of states which exhibits inequivalence
is of practical importance or, instead, is limited to such a tiny
range of forces which would make it vanishing for the sake of every
operational purpose. Besides checking the validity of the analytical
approach, the computational analysis described in the next two
sections will also try to address this problem, by investigating
the transition from the non-equivalence to the equivalence regimes.

\section{Simulation Details}
The investigations on the ensemble difference have been carried out by
simulating the Langevin dynamics of two different chain models by
sampling the isometric and in the isotensional ensembles for different
chain lengths, using the ESPResSo simulation
package\cite{limbach06a}. Each chain consists of a given number of
monomers, which are interacting only with their respective first
neighbors via an harmonic potential. Therefore no excluded volume
interactions are present and in the free case the chains perform a
pure random walk.  The role of Langevin equation
\begin{equation}
        \label{langevin}
          \frac{d^{2}\mathbf{x}_{i}}{d t^{2}} = \mathbf{F}_{i} - \gamma \frac{d \mathbf{x}_{i}}{dt} + \mathbf{W}_i(t) \\
\end{equation}
is basically that of providing a thermostat for the chain, where the
position of the monomer $i$ is $\mathbf{x}_i$, $\mathbf{F}_i$
represent the conservative force acting on the monomer.  As usual, the
thermostat is acting via a friction coefficient $\gamma$, and a random
force $\mathbf{W}_i(t)$ with zero mean and square deviation $\langle
\mathbf{W}_i(t)\cdot\mathbf{W}_j(t^\prime) \rangle= 6 k_{B} T \gamma
\delta_{ij} \delta(t-t^\prime)$, in order to satisfy the
fluctuation-dissipation theorem.

By measuring energies in units of $k_BT$, distances in arbitrary
units $ d $, and considering unitary masses it is possible to express the energy of a chain
consisting of $N$ monomeric units, depicted in
Fig.\ref{gauss_chain_ensembles}, as
\begin{eqnarray}
     \label{harmonic} 
        U=\frac{1}{2} k \sum_{i=1}^N ( | {\bf x}_{i}-{\bf x}_{i-1}|-r_{eq})^{2},
\end{eqnarray}
where $k = 20/  d ^{2} $ is the spring constant and $r_{eq}$ is the equilibrium
distance between a pair of connected monomers.

We focused our attention on two specific cases, namely that of $r_{eq}
=  d $ and that of zero equilibrium distance. In particular the latter case
models precisely a Gaussian probability distribution for the end-to-end
distance, in the free case.  For every simulation, the friction
coefficient was set to $\gamma=0.5/\tau$ and the integration time step was
set to 0.01 $\tau$, where $\tau =  d $ is the characteristic time.

The sampling of the isometric ensemble was realized by fixing the
spatial position of both terminal monomers, $\mathrm{x}_1$ and
$\mathrm{x}_N$. In the isotensional ensemble, the end monomer of the
chain was fixed, while the other monomers one was free to move and a
given force, constant in modulus and direction, was applied on it.

The simulation procedures employed for the isometric ensemble
consisted of constraining the positions of the first and last monomers
at the desired distance $X$, then performing an relaxation run, up to
at least 10 times the auto-correlation time of the observable of interest, $F$ and
eventually computing the average $ \langle F \rangle= \langle \FF \cdot \XX / X
\rangle $, where $\XX = \mathbf{x}_N - \mathbf{x}_1$.  In the isotensional ensemble case,
the chains started from a straight
conformation where the position of the first monomer was constrained,
and a constant force $\FF$ was applied to the last monomer. Then,
after relaxing the system, the end-to-end distance $X$ was sampled.
By varying over suitable ranges the end-to-end distance in the
isometric ensemble and the magnitude of the applied force in the
isotensional ensemble, we sampled the force-extension curves $\langle
F\rangle(X)$ and $F(\langle X\rangle)$ for a number of different chain
lengths ranging from 6 to 500 monomeric units. Every point in the force-extension curves
was generated form the average taken during a $10^8$ steps long run.

\section{Results and Discussions}
\begin{figure}[ptb]
     \includegraphics[width=\columnwidth]{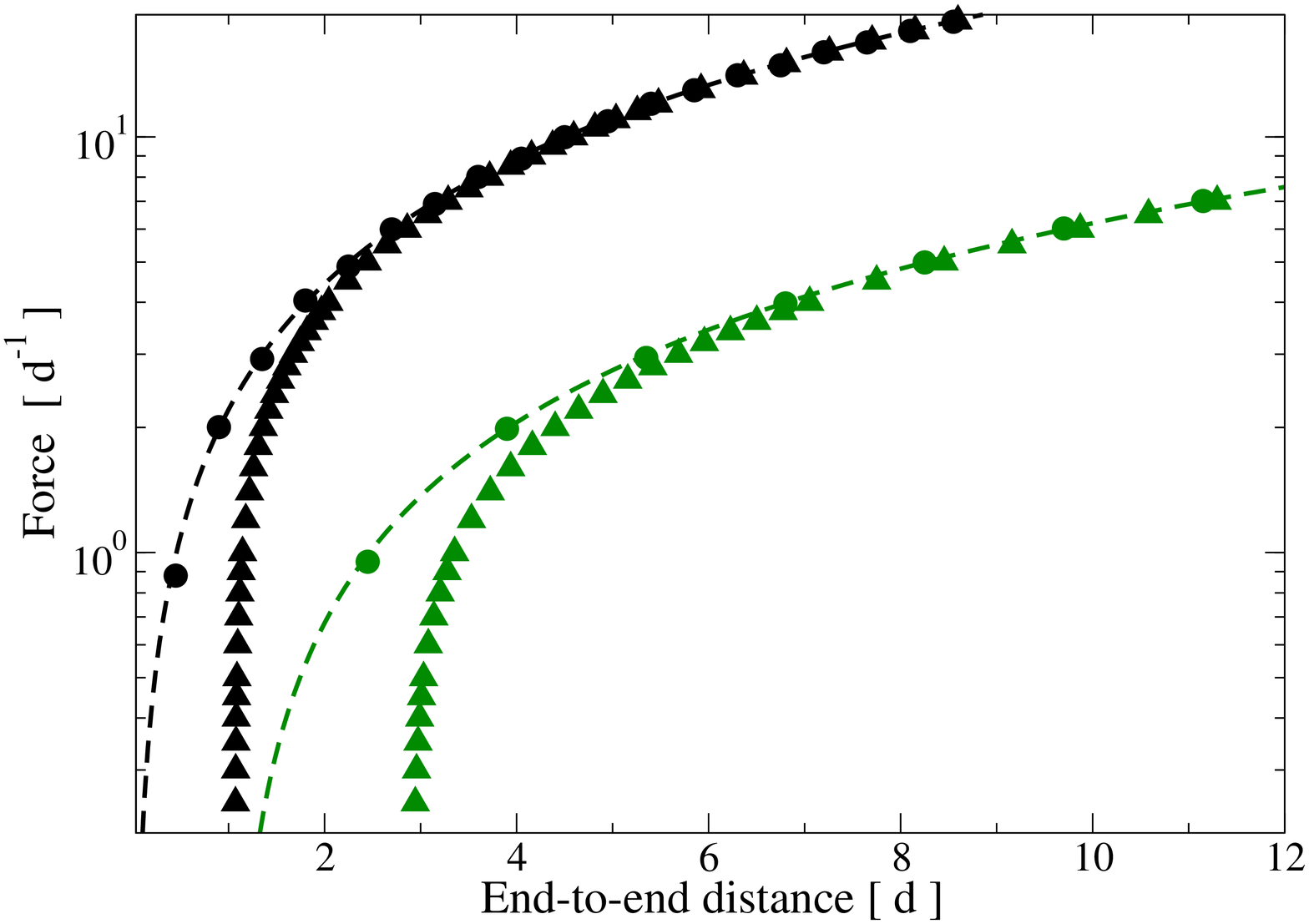}
    \caption{The force-extension curve for the $r_{eq}=0$ case in the isometric (circles)
    and isotensional (triangles) ensembles. From left to right, $N=10$ (black), $N=30$
    (green). The dashed lines show the result of linear interpolation. Data has been offset
    along the x-axis for the sake of clarity.\label{fec_unscaled_req0}}
\end{figure}

\begin{figure}[ptb]
     \includegraphics[width=\columnwidth]{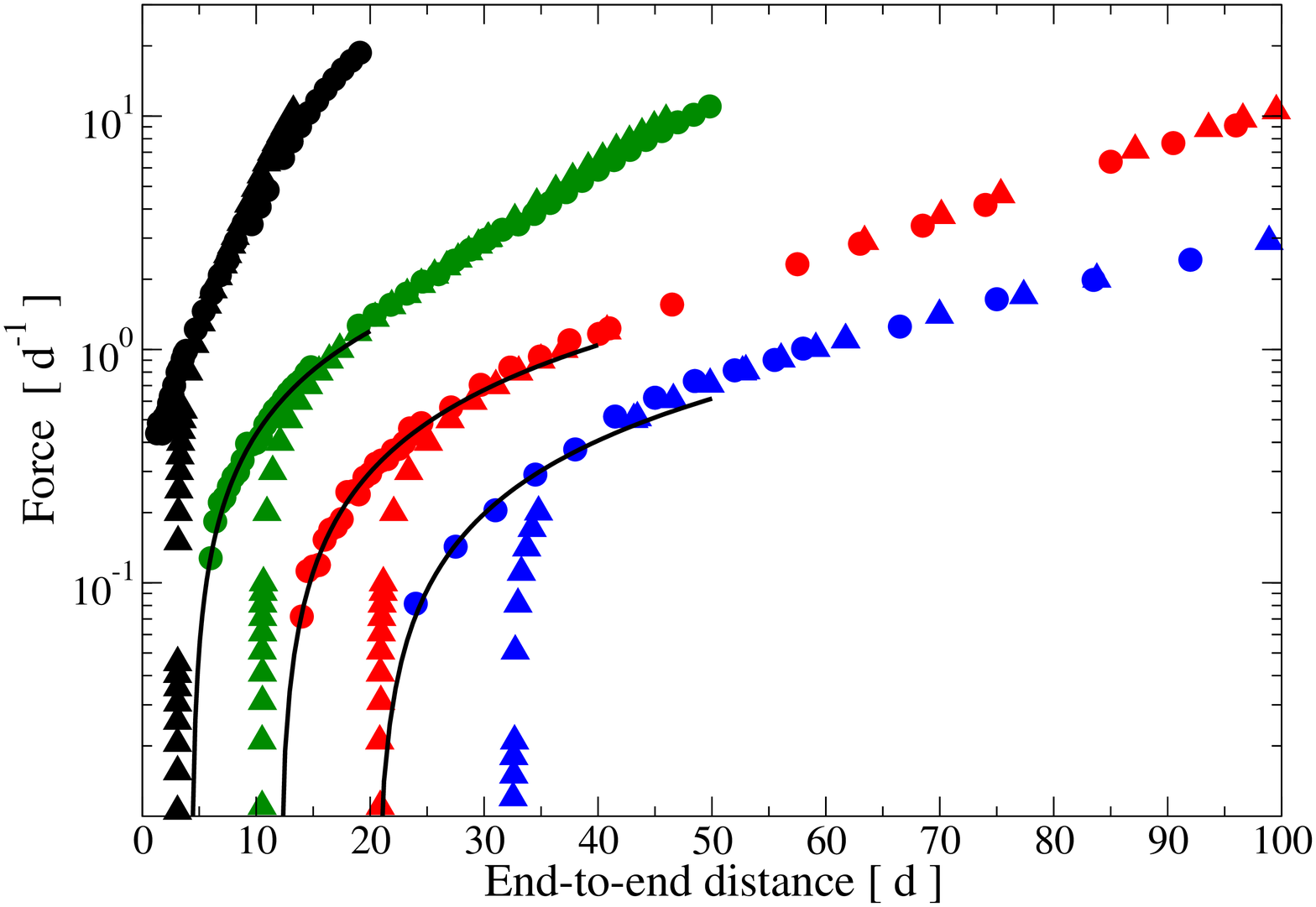}
     \caption{The force-extension curve for the $r_{eq}= d $ case in the
       isometric (circles) and isotensional (triangles)
       ensembles. From left to right, $N=10$ (black), $N=30$ (green),
       $N=60$ (red), $N=90$ (blue).The solid lines represent the
       result of linear interpolation in the low-stretching regime.
       Data has been offset along the x-axis for the sake of
       clarity.\label{fec_unscaled_req1}} 
\end{figure}

In Fig.(\ref{fec_unscaled_req0}) and (\ref{fec_unscaled_req1}) we
present the measured force-extension curves for the $r_{eq}=0$ and
$r_{eq}= d $ cases, respectively.  From the qualitative point of view,
every force-extension curve displays the same pattern, namely a
difference $\Delta^\prime(X)$ which decreases with increasing
applied (or measured) force. This behavior is somewhat expected, and
can be qualitatively explained in the following way. In the
isotensional ensemble, when the external force is vanishing, $\langle
X\rangle$ is expected to have a non-zero average value $\langle
X\rangle_0$, the exact value of which depends on $r_{eq}$.  In the
isometric case, on the other hand, although the projected force
$\langle F\rangle$ is not strictly defined at zero end-to-end
distance, simple symmetry arguments show that the force acting on each
of the terminal beads has, on average, to be zero. Therefore, a
vanishing value of $\langle F\rangle$ is expected in this limit. While
this arguments account for the differences in the weak stretching
regime, in the high stretching regime the energetic contributions to
the free energy are expected to dominate over the entropic ones, and
therefore a narrowing of the distance between the force-extension
curves is expected. In Fig.(\ref{fec_unscaled_req1}) the result of a
linear fit on the whole spanned x-range is also included,
showing the perfect Hooke behavior of chain in the isometric ensemble.
In contrast to the $r_{eq}=0$ case, the response for the model with a
non-zero equilibrium distance does not display a linear behavior over the
whole end-to-end distance range. This feature appears because in this case two
characteristic distances enter the description of the model, namely,
the root mean square displacement of a bead around the equilibrium
position, and the equilibrium distance itself. Indeed, since the
spring constant $k$ is relatively high with respect to the thermal
energy, at low applied forces (or, equivalently, at short end-to-end
distances) the chain behaves much like a freely-jointed chain, while
when the applied force increases and the springs are significantly
stretched with respect to $r_{eq}$, a different, effective bond length
becomes relevant.

\begin{figure}[ptb]
\includegraphics[width=\columnwidth]{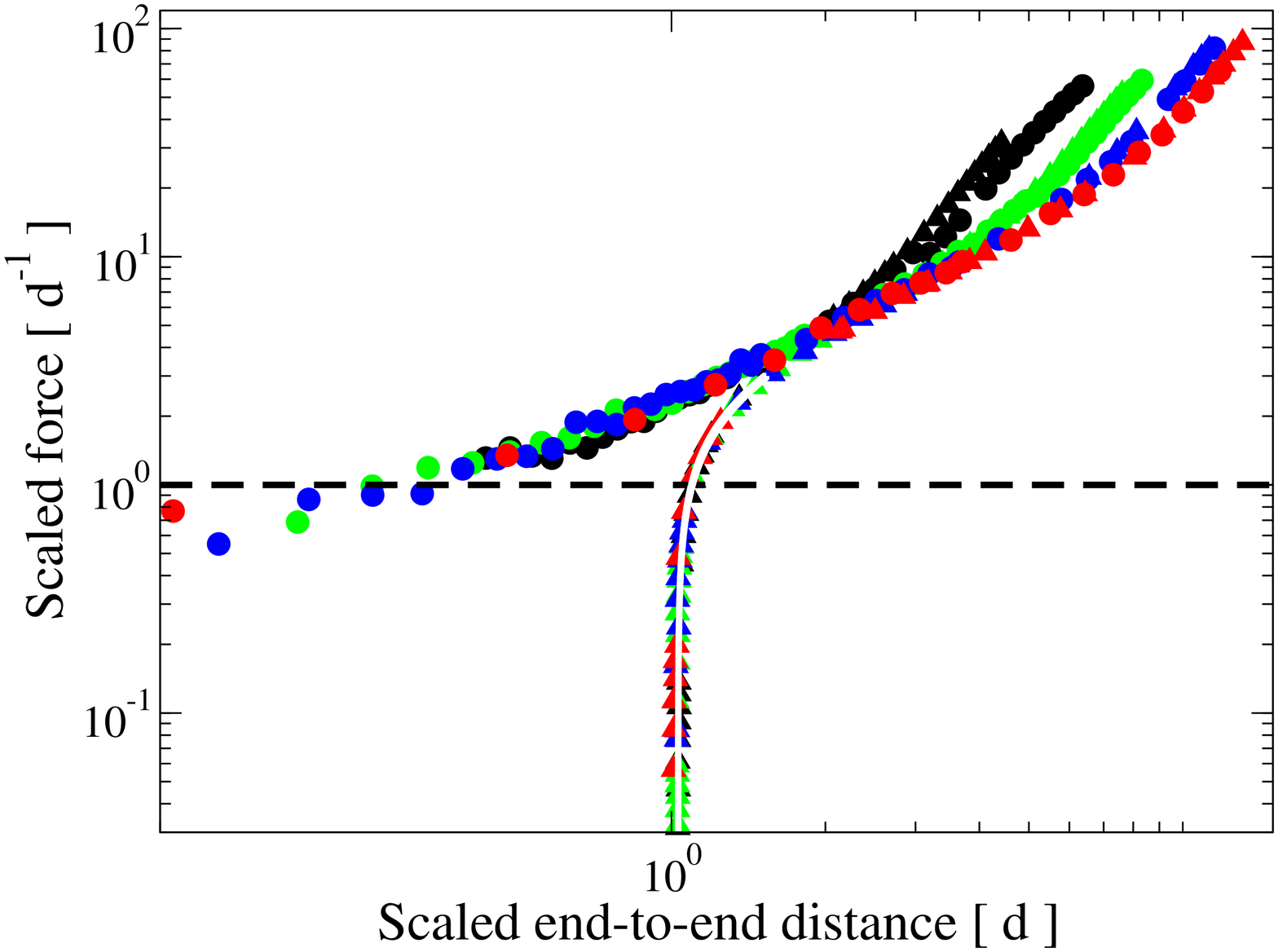}
\caption{ Universality of the force-extension curves in the
  low-stretching regimes for the $r_{eq}= d $ case, reporting the
  rescaled force $\widetilde{F}$ as a function of the rescaled
  end-to-end distance $\widetilde{X}$ for the isometric (circles) and
  isotensional (triangles) ensembles.  The white solid line on top of
  the isotensional sampled curve in the low-stretching regime is a fit
  to Eq.(\ref{fluct_x}), while the horizontal dashed line represents
  an estimate for the low-stretching regime from dimensional analysis.
\label{fec_scaled} }
\end{figure}

The definition of a weak and a high stretching region is of course
somewhat arbitrary, although by dimensional analysis, the obvious
threshold is set by $F^{c} = k_BT / a $, where $a$ now identifies the
model-dependent effective bond-length. This can be
defined\cite{doi86a} according to the scaling behavior of a linear
chain in the free case, $a^2\equiv \langle\XX^2\rangle_0 /
(N-1)$. Since in the weak stretching regime the force is expected to
be a small perturbation with respect to the thermal fluctuations, the
question naturally arises, whether and to which extent do the scaling
arguments hold in the isometric and isotensional ensemble that are
valid for the free case only. In order to check this, the
force-extension curves have to be rescaled. The end-to-end distance
behavior is obviously $X\sim\sqrt{N-1}$, while from Eq.(\ref{FEC_F}), and
remembering that in a Gaussian chain $b \sim 1/\sqrt{N-1}$, one can
obtain the scaling behavior for the force $F \sim 1/\sqrt{N-1}$.  By
plotting the rescaled end-to-end distance and forces, namely,
\begin{eqnarray}
\widetilde{X}&=&X/\sqrt{N-1}\nonumber \\
\widetilde{F}&=&F \sqrt{N-1}\nonumber
\end{eqnarray}
for every different chain length, one can see (Fig. \ref{fec_scaled})
that in the low stretching regime both the isotensional and isometric
force-extension curves collapse onto the same universal curve. The
force-extension curves in the two ensembles fall onto two different
universal curves only up to a certain point, after which they start to
depart from the universal behavior of free chains and, at the same
time, the difference between the isometric and isotensional cases (for
a given chain length) diminishes dramatically. In fact, as long as the
scaling laws of free chains are fulfilled, they retain the
characteristic differences in the force-extension curves. On the other
hand, when the chains start being moderately or strongly stretched,
the scaling law changes, and the ensemble difference becomes less
pronounced.
The universal behavior of the chains in the low-stretching regime
already points out that the ensemble difference $\Delta$ should become
constant in this regime, since both $\Delta^\prime(X)$ and $X^*$
scale like $\sqrt{N-1}$. The only requirement for the loss of
ensemble equivalence is that the scaled force $\widetilde{F}$ is
roughly less than 1.0. This implies that, for example, in order to
keep a constant ensemble difference, the applied forces should
decrease as $1/\sqrt{N-1}$ with increasing chain length.

\begin{figure}[ptb]
\includegraphics[width=\columnwidth]{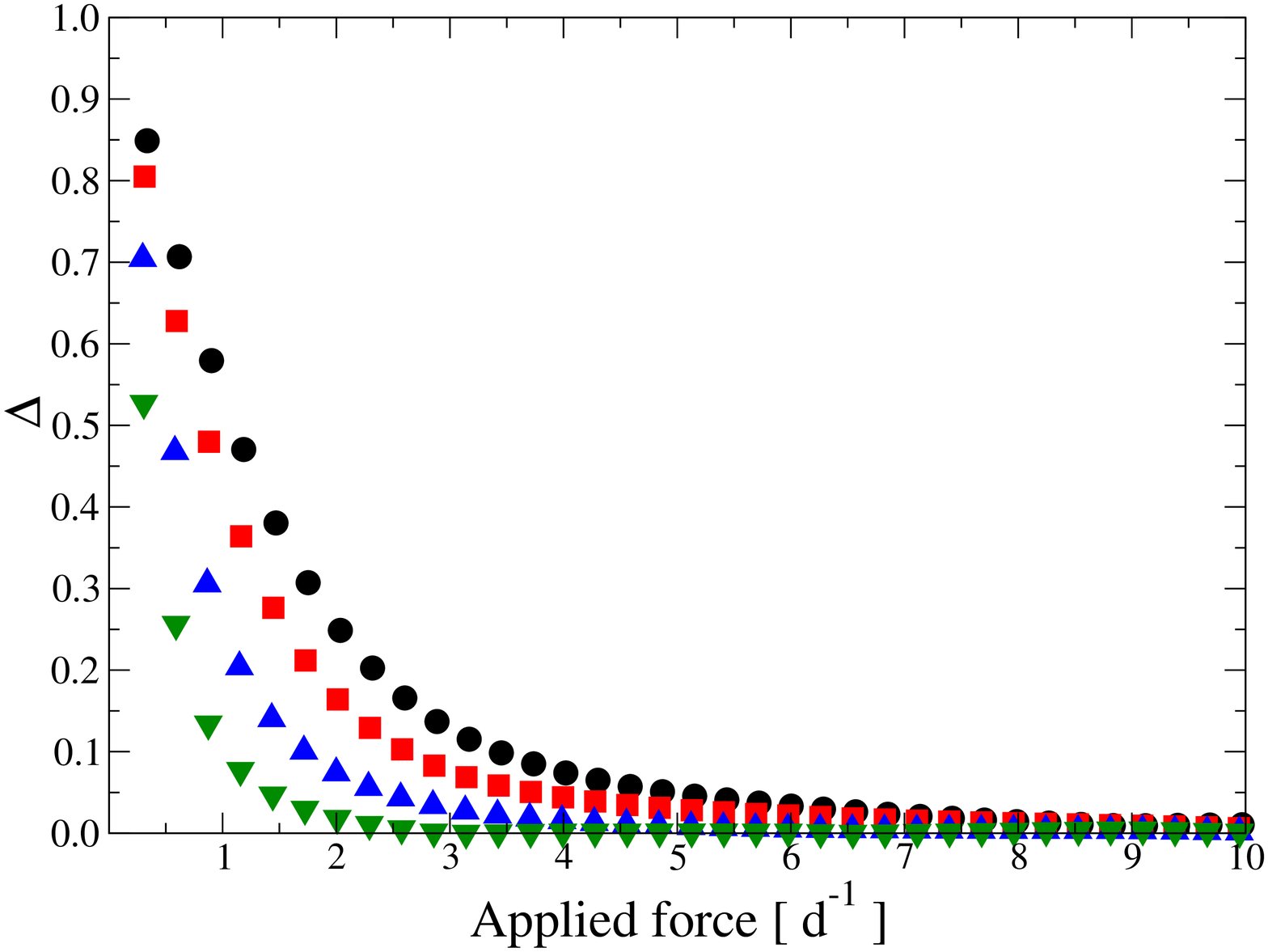}
\caption{ Dependence of the ensemble difference $\Delta$ as a function
  of the force applied in the isotensional ensemble, for different
  chain lengths, $N=15$ (circles), $N=25$ (squares), $N=60$ (upper
  triangles) and $N=170$ (lower triangles) in the $r_{eq}=0$
  case.\label{ens_diff}}
\end{figure}

Although the scaling arguments are helpful in showing the ensemble
inequivalence in the low-stretching regime, it is instructive to
explicitly look at the dependence of the ensemble difference on the
force, for different chain lengths and, also, to look for the scaling
behavior of $\Delta$ in different force regimes.  In
Fig.\ref{ens_diff}, the ensemble difference is shown, as a function of
the applied force, for different chain lengths for the $r_{eq}=0$
case.  As it has been already noticed from the qualitative analysis of
the force-extension curves, the stronger the applied force, the
smaller is the ensemble difference.  The decay to zero of the ensemble
difference happens for smaller forces, the longer the chain is, thus
showing that in the moderate stretching regime weaker forces are
required to reach the ensemble equivalence. On the other hand, the
behavior of the ensemble difference in the weak stretching regime
suggests that all the curves tend to converge to the limiting value of
1, no matter what the system size is.

\begin{figure}[ptb]
\includegraphics[width=\columnwidth]{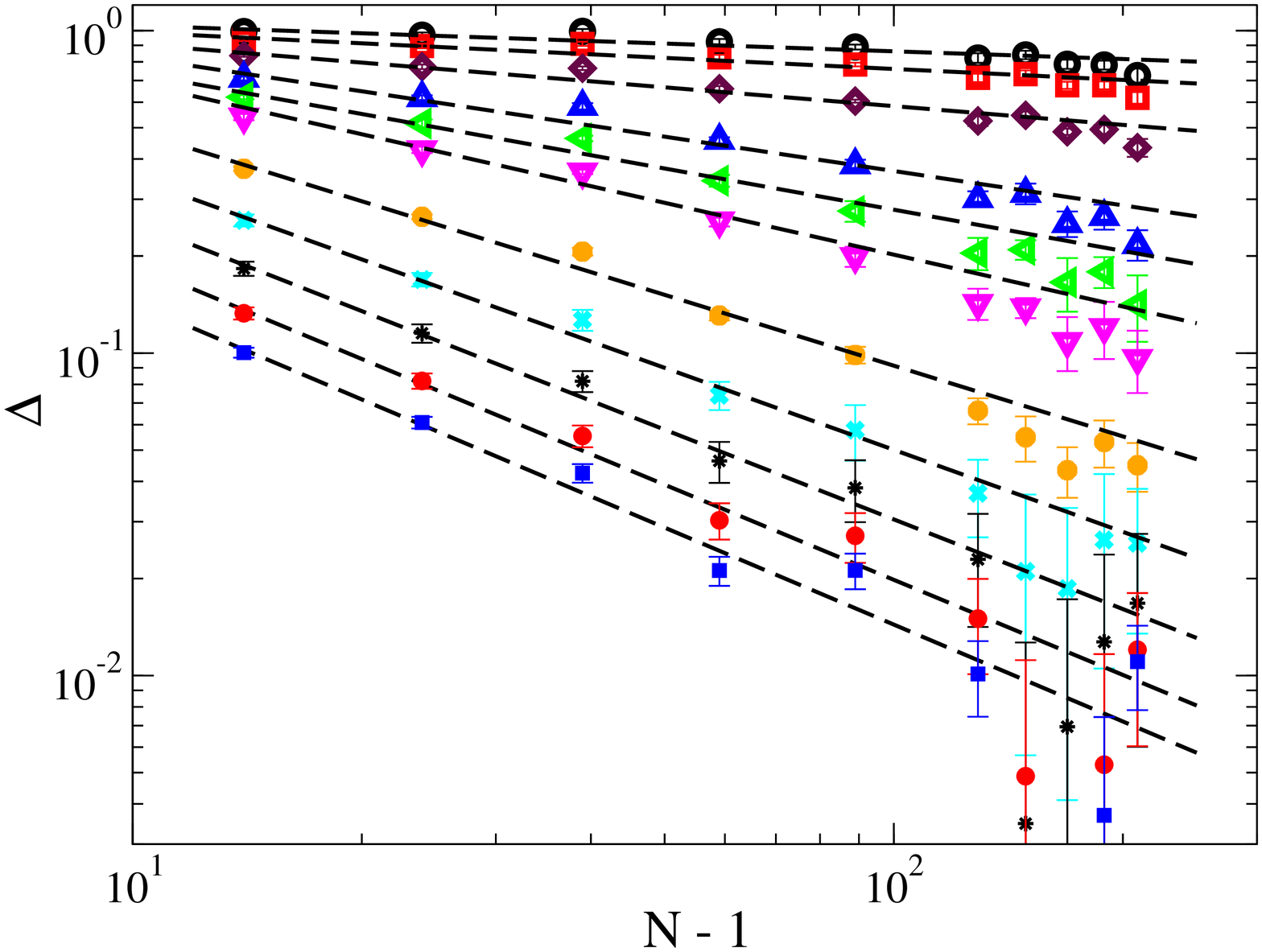}
\caption{Logarithmic plot of the ensemble difference $\Delta$ as a
  function of the number of beads, $N-1$, for the $r_{eq}=0$
  case. Different symbols correspond to different applied forces in
  the isotensional ensemble, going from the low-stretching regime
  (smaller slope) to the high-stretching regime (larger slope). Dashed
  lines correspond to the best fit to a power law. The applied forces
  ranged from 0.035 to 3.25 $d^{-1}$ \label{ens_diff2}}
\end{figure}

\begin{figure}[ptb]
\includegraphics[width=\columnwidth]{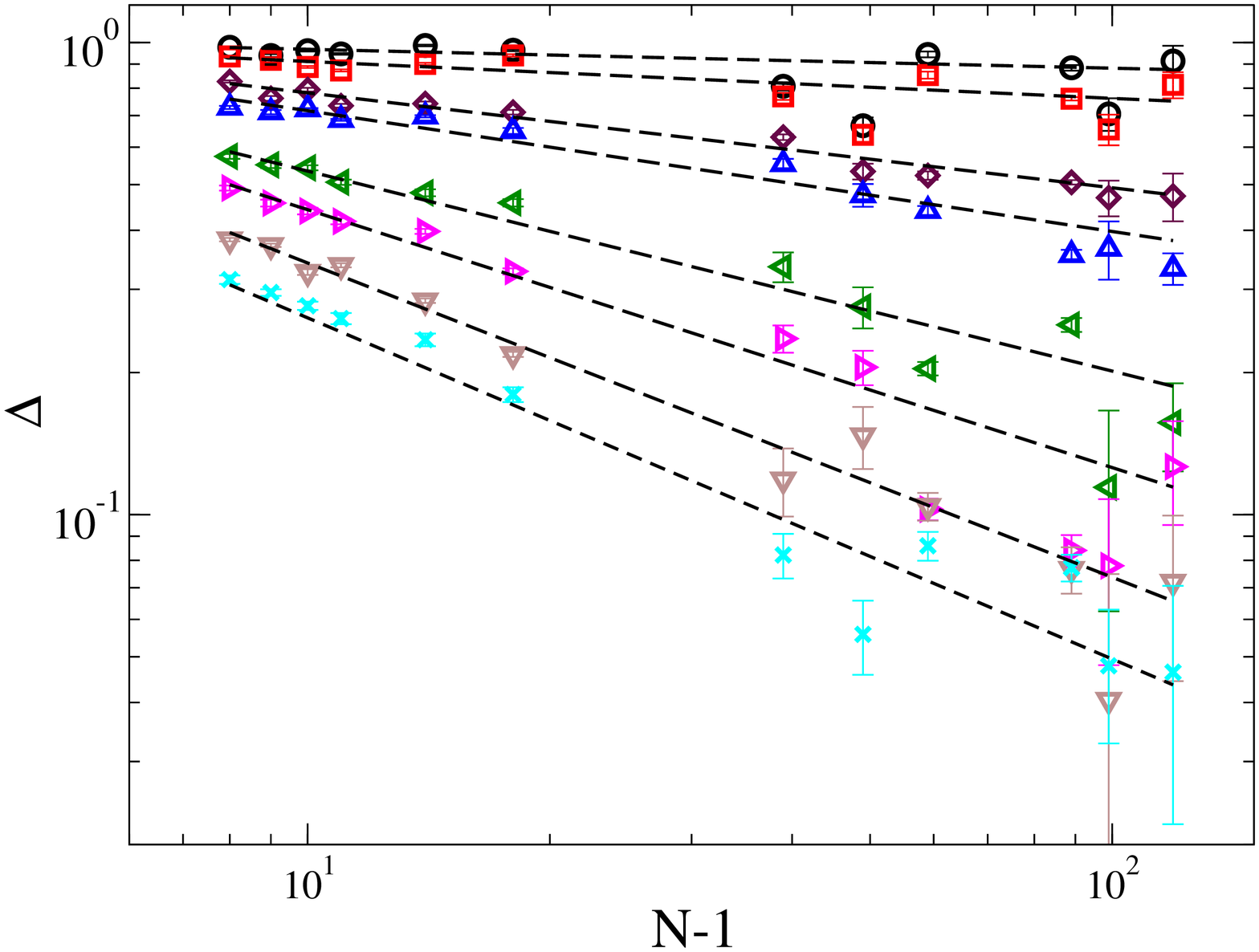}
\caption{Logarithmic plot of the ensemble difference $\Delta$ as a
  function of the number of beads, $N-1$, for the $r_{eq}= d $
  case. Different symbols correspond to different applied forces in
  the isotensional ensemble, going from the lie-stretching regime
  (smaller slope) to the high-stretching regime (larger slope). Dashed
  lines correspond to the best fit to a power law. The applied forces
  ranged from 0.035 to 0.7 $d^{-1}$ \label{ens_diff3}}
\end{figure}

If one focuses instead of the force dependence on the dependence of
the number of monomers on the ensemble
difference, it is
possible to investigate the scaling properties of the ensemble
difference in different force regimes. In this way we perform a finite-size
study of the convergence of the isometric and isotensional
ensembles. In Fig.\ref{ens_diff2} and \ref{ens_diff3} the scaling of
$\Delta$ with respect to the number of units in the chains is shown
for the $r_{eq}=0$ and $r_{eq}= d $ cases, respectively.
\begin{figure}[ptb]
\includegraphics[width=\columnwidth]{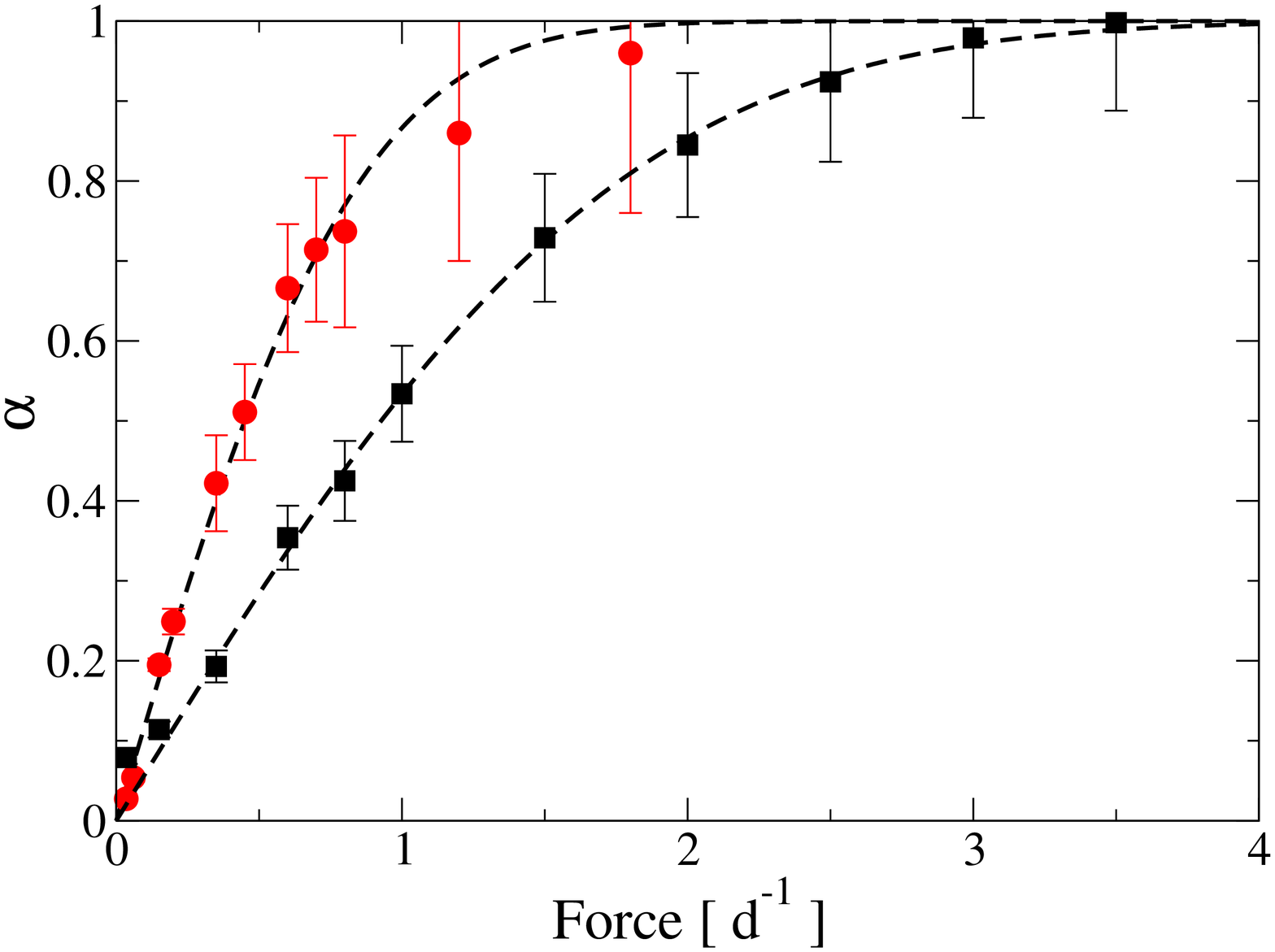}
\caption{Dependence of the fitted scaling exponent $\alpha$ on the
  applied force for the $r_{eq}=0$ (squares) and $r_{eq}= d $ (circle)
  cases.  The dashed lines are the result of a best fit to the error
  function.
\label{exponent}}
\end{figure}

For both models we observe the same trend, and for each value of the
force the ensemble difference actually follows a power-law $\Delta(N) =
A (N-1)^{-\alpha}$ (the result of a best fit being also shown in the
plots), where the actual value of the exponent strongly depends on the
value of the force itself. For diminishing values of the applied
force, the curves in the logarithmic plot show a decreasing slope,
displaying the clear tendency to become a constant (namely, one) in
the limit of vanishing forces. This result not only confirms the
inequivalence of the isometric and isotensional ensembles in the
vanishing force limit, but also shows the development of the scaling
behavior in the intermediate force regime. The limiting behavior in
the high-stretching regime can be derived by evaluating the ensemble
difference using Eq.(\ref{fluct_x}) and (\ref{FEC_X}), and looking for
the asymptotic behavior at large values of $\nu$, leading to $ \Delta
\sim (N-1)^{-1} $ in the limit of large $\nu$.  The two limiting
behaviors of the ensemble difference are obviously characterized by
the ratio of the typical energies with the thermal one, $\beta F/b$
which is vanishing and greater then one in the free and overstretched
limits, respectively. The way the overstretched region is approached,
in terms of scaling exponents will therefore depend on the typical
length $1/b$ associated with the model. In Fig.(\ref{exponent}) the
value of the scaling exponent of $\Delta$ as a function of the applied
force is shown for the two models. Indeed, the high-stretching
exponent $\alpha=1$ is reached much faster in the $r_{eq}=1$ case.  A
phenomenological fitting function is also reported, which describes
surprisingly well the switch between the two regimes is $\alpha(F) =
\mathrm{erf}(cF)$, where $c\simeq 0.5$ and $1.0$ for $r_{eq}=0 $ and
$r_{eq}= d $, respectively (notice that the effective bond length $a$ 
for the two cases are roughly 0.35 and 1.03). 
The important information that is conveyed from this analysis, as
anticipated  in Sec.\ref{quantens}, is that the transition from the inequivalence to the equivalence regime actually encompasses a surprisingly broad range of forces, being the scaling exponent still sensibly different from one at reduced forces of about 1 and 2 for the $r_{eq}=1$ and $r_{eq}=0$ cases, respectively.
The scenario of a scaling exponent which would reach unity for values of reduced force much smaller than one, has therefore been ruled out. This result thus demonstrates that the ensemble inequivalence has actually an important impact from the operational point of view on measurements in single molecule experiments.

\section{Conclusions}
Computer simulations have been employed to study the behavior of
linear model polymers in two different ensembles, namely, the
isotensional and isometric one, which are representative for a class
of single molecule experiments. In particular, we have addressed the
question of the equivalence of these two ensembles in the
thermodynamic limit of growing chain lengths.  Finite size effects of
order $(N-1)^{-1}$ for the ensemble difference have been demonstrated
as a limiting case for a Gaussian chain experiencing high tensions,
and verified for bead-spring models with zero and non-zero
equilibrium distance. In this case ensemble equivalence is reached in
the infinite chain length limit. By switching to the low-stretching
regime, a dramatic change in the scaling behavior appears. Here we
find that the force-extension curves exhibit an universal scaling
behavior that is typical for free Gaussian chains in
equilibrium. This, in turn, leads to the fact that ensemble
equivalence can indeed never be obtained in the vanishing force
limit, as has been pointed out by Neumann \cite{neumann03a}.  Our
computer simulations confirm and enhance (by showing the relevance
of the inequivalence to equivalence regime transition) the analysis
by Neumann that care has to be taken when considering the thermodynamics
of single molecules, which presents many subtle differences with
respect to bulk systems, despite the formal analogies between the
two.

\acknowledgments 
We thank D. Keller, R. M. Neumann and M. Rubi for correspondence and
helpful remarks, and the Frankfurt Center for Scientific Computing for
allocating the computer time necessary for this project.  Financial
support by the Volkswagen foundation is gratefully acknowledged.
 
\bibliographystyle{apsrev}
\bibliography{ensemble_inequivalence_suzen}
\newpage
\end{document}